\documentstyle[12pt]{article}

\newcommand{\postscript}[2]{\vspace{5cm}}
\newcommand{\postbb}[3]{\vspace{5cm}}

\newcommand{\beq}{\begin{equation}}
\newcommand{\eeq}{\end{equation}}
\newcommand{\beqa}{\begin{eqnarray}}
\newcommand{\eeqa}{\end{eqnarray}}
\def\ra{\rightarrow}
\def\RA{\rightarrow}
\def\etal{{\it et al.}}
\newcommand{\RMP}[3]{{\em Rev. Mod. Phys.} {\bf #1}, #2 (19#3)}
\newcommand{\PR}[3]{{\em Phys. Rev.} {\bf #1}, #2 (19#3)}
\newcommand{\PL}[3]{{\em Phys. Lett.} {\bf #1}, #2 (19#3)}

\newcommand{\PRL}[3]{{\em Phys. Rev. Lett.} {\bf #1}, #2 (19#3)}
\newcommand{\NP}[3]{{\em Nucl. Phys.} {\bf #1}, #2 (19#3)}
\newcommand{\con}[3]{{\bf #1}, #2 (19#3)}

\begin{document}
\begin{flushright} NSF-ITP-95-140, UPR-0683T  \\
October 31, 1995 \end{flushright}

\vspace{1.5cm} 
\centerline{\large \bf
Precision Experiments, Grand Unification,
and Compositeness\footnote{Invited talk presented at
{\it SUSY-95}, Palaiseau, France,
May 1995. The results presented here are from
a collaboration with Jens Erler. For more details,
see~\cite{pdg95}.}}
\vspace*{6.0ex}
\centerline{\large Paul Langacker}
\vspace*{1.5ex}
\centerline{\large \it
University of Pennsylvania,
Department of Physics}
\centerline{\large \it
Philadelphia, Pennsylvania, USA 19104-6396}
\vspace*{1.5ex}
\centerline{\it and}
\vspace*{1.5ex}
\centerline{\large \it
Institute for Theoretical Physics,}
\centerline{\large \it University of California, Santa Barbara, CA 93106-4030}

\vspace*{4.5ex}
\begin{abstract}
Precision electroweak data allow one to test the standard model,
constrain its parameters, and search for the effects of some kinds
of new physics. The results of the most recent data from LEP, SLC, and
elsewhere are described, as are their predictions for
$m_t$, $M_H$, and $\alpha_s$.
The implications for
the two major classes of extensions of the standard model,
supersymmetry/unification and compositeness, are described.
\end{abstract}

\section{Recent Data}

The four LEP experiments ALEPH,
DELPHI, L3, and OPAL have recently presented combined results
at the Brussels Europhysics Conference and at the Lepton-Photon
conference in Beijing. These include preliminary results from
the  1994 run, and altogether include nearly 14 million events.
The averages include
a proper treatment of common systematic uncertainties~\cite{a1}.
Similarly, the
SLD experiment at the SLC~\cite{a2} has presented new results on the
left-right asymmetry $A_{LR}$, including the data from the very successful
1994-1995 run, as well as the first direct determination of final state
couplings
of the $b$ and $c$.
The major $Z$-pole results are shown in Table~\ref{tab1}.
The first row
gives the value of the $Z$ mass, which is now known to remarkable
precision.  Also shown are the lineshape variables $\Gamma_Z$, $R$, and
$\sigma_{\rm had}$; the heavy quark production rates; various
forward-backward asymmetries, $A_{FB}$; quantities derived from the $\tau$
polarization $P_{\tau}$ and its angular distribution;
final state $b$ and $c$ couplings determined by SLD from mixed
polarization forward-backward asymmetries; and the effective
weak angle $\bar{s}^2_\ell$ obtained from the jet charge asymmetry. $N_\nu$
is the number of effective active neutrino flavors with masses light enough
to be produced in $Z$ decays.  It is obtained by subtracting the widths for
decays into hadrons and charged leptons from the total width $\Gamma_Z$
from the lineshape.  The asymmetries are expressed in terms of the
quantity
\beq A^o_f = \frac{2 \bar{g}_{Vf} \; \bar{g}_{Af}}{\bar{g}^2_{Vf}
+ \bar{g}^2_{Af}},  \label{eqn1} \eeq
\newpage
\noindent
where $\bar{g}_{V,Af}$ are the vector and axial vector couplings to fermion
$f$.

\begin{table} \centering 
\begin{tabular}{|lcc|}
\hline 
Quantity & Value & Standard Model \\ \hline
$M_Z$ (GeV) & $91.1884 \pm 0.0022$ &  input \\
$\Gamma_Z$ (GeV) & $2.4963 \pm 0.0032$ & $2.497 \pm 0.001 \pm
0.002 \pm [0.002] $ \\
$R = \Gamma({\rm had})/\Gamma(\ell \bar{\ell})$ & $20.788 \pm
0.032$ & $20.77 \pm 0.004 \pm 0.002 \pm [0.02]$ \\
$\sigma_{\rm had} = \frac{12 \pi}{M_Z^2} \; \frac{\Gamma(e
\bar{e}) \Gamma({\rm had})}{\Gamma_Z^2} ({\rm nb})$ & $41.488 \pm 0.078$
& $41.45 \pm 0.002 \pm 0.004 \pm [0.02]$ \\
$R_b = \Gamma(b \bar{b})/ \Gamma({\rm had})$ &$0.2219 \pm 0.0017$
& $0.2156 \pm 0 \pm 0.0003$ \\
$R_c = \Gamma(c\bar{c}) /\Gamma({\rm had})$ & $0.1540 \pm 0.0074$
 & $0.172 \pm 0 \pm 0$ \\
$A^{0\ell}_{FB} = \frac{3}{4} \left( A_{\ell}^0 \right)^2$ &
$0.0172 \pm 0.0012$ & $0.0155 \pm 0.0004 \pm 0.0004$  \\
$A^0_{\tau} \left(P_\tau \right)$ & $0.1418 \pm 0.0075$ & $0.144
\pm 0.002 \pm 0.002$ \\
$A^0_e \left( P_\tau\right)$ & $0.1390 \pm 0.0089$ & $0.144 \pm
0.002 \pm 0.002$ \\
$A^{0b}_{FB} = \frac{3}{4} A^0_e A^0_b$ & $0.0997 \pm 0.0031$ &
$0.101 \pm 0.001 \pm 0.001$ \\
$A^{0c}_{FB} = \frac{3}{4} A^0_e A^0_c$ & $0.0729 \pm 0.0058$ &
$0.072 \pm 0.001 \pm 0.001$ \\
$\bar{s}^2_{\ell} \left(A^{Q}_{FB} \right)$ & $0.2325 \pm
0.0013$ & $0.2319 \pm 0.0002 \pm 0.0002$ \\
$A^0_e \left(A^0_{LR} \right)$ \ \ (SLD) & $0.1551 \pm 0.0040 $
& $0.144 \pm 0.002 \pm 0.002$ \\
$A^0_b $ \ \ (SLD) & $0.841 \pm 0.053 $ & $0.934 \pm 0 \pm 0$ \\
$A^0_c $ \ \ (SLD) & $0.606 \pm 0.090 $ & $0.667 \pm 0.001 \pm 0.001$ \\
$N_\nu$ & $2.991 \pm 0.016$ & $3$ \\ \hline
\end{tabular}
\caption{$Z$-pole observables from LEP and SLD compared to their standard
model expectations.  The standard model prediction is based on $M_Z$ and
uses the global best fit values for $m_t$ and $\alpha_s$, with $M_H$ in the
range $60 - 1000$~GeV.}
\label{tab1}
\end{table}

{}From the $Z$ mass one can predict the other observables,
including electroweak loop effects.  The predictions also depend on the top
quark and Higgs mass, and $\alpha_s$ is needed for the QCD corrections to
the hadronic widths.  The predictions are shown in the third column of Table~1,
using the value $m_t = 180 \pm 7 $~GeV obtained for $M_H = 300$~GeV in
a global best fit to all data (including the direct determination
$m_t = 180 \pm 12$ GeV by CDF~\cite{cdfmt} and DO~\cite{d0mt}).
The first uncertainty is from $M_Z$ and
$\Delta r$ (related to the running\footnote{There have been several recent
reevaluations~\cite{je}-\cite{ms}
of the hadronic contribution to the running of
$\alpha$. Following a correction to~\cite{ms} these are in reasonable
agreement.
We use $\alpha(M_Z)^{-1} = 128.09 \pm 0.09$ from~\cite{je}.}
of $\alpha$ up to $M_Z$), while the
second is from $m_t$ and $M_H$, allowing the Higgs mass to vary in the
range $60 - 1000$~GeV.  The last uncertainty is the QCD uncertainty from
the value
of $\alpha_s$.  Here the value and uncertainty are given by
$\alpha_s = 0.123 \pm 0.004$, obtained from the global fit to the lineshape.

The data is generally
in excellent agreement with the standard model predictions.
However,
\beq R_b = \frac{\Gamma(b\bar{b})}{\Gamma(\rm had)} = 0.2219 \pm 0.0017 \eeq
is $3.7\sigma$ higher than the standard model expectation,
while $R_c = \Gamma(c\bar{c})/\Gamma(\rm had) $
is 2.4$\sigma$ below. These are correlated: if $R_c$ is
fixed at the standard model value of 0.172, then~\cite{a1}
$R_b = 0.2205 \pm 0.0016$, which is still 3.0$\sigma$ too high.
Within the standard model framework, these must be considered
statistical fluctuations or systematic errors. However,
because of special vertex corrections, the $b
\bar{b}$ width actually decreases with $m_t$, as opposed to the other
widths which all increase.  This can be seen in Figure~\ref{fig1}.
Thus, $R_b$ favors low values for $m_t$. By itself $R_b$ is insensitive
to the Higgs mass $M_H$, but when combined with
other observables, for which the $t$ quark and Higgs mass $M_H$ are strongly
correlated, $R_b$ favors low values for $M_H$.
Another possibility, if the effect is more than a
statistical fluctuation, is that it may be due to some sort of new physics.
Many types of new physics will couple preferentially to the third
generation, so this is a serious possibility.
As will be seen below, the possibility of new physics in the $Z b \bar{b}$
vertex is strongly correlated with the value of $\alpha_s$ extracted
from the $Z$ lineshape.

\begin{figure}
\postbb{40 220 530 680}{/home/pgl/fort/nc/graph/gam/xxrb.ps}{0.6}
\caption{ Standard model prediction for $R_b \equiv \Gamma(b \bar{b}) /
\Gamma( \rm had)$ as a function of $m_t$, compared with the LEP
experimental value.  Also shown is the
range $180 \pm 12$~GeV determined by the direct CDF and D0 observations.}
\label{fig1}
\end{figure}

Another discrepancy is the value of the left-right asymmetry
\beq A^0_{LR} = A^0_e = 0.1551 \pm 0.0040
\eeq
obtained by the SLD collaboration using all data from 1992-1995.
This value has moved closer to the standard model expectation of
$0.144 \pm 0.003$ than the  previous value of $0.1637 \pm 0.0075$
from 1992-1993. However, because of the smaller error it is still
some $2.3\sigma$ higher than
the standard model prediction.  This result (combined with $M_Z$)
favors a large value of the top quark mass, around 220~GeV, which
is not in good agreement with other observables.  One possibility is that
it is pointing to new physics.  Possibilities here would include $S < 0$,
where $S$ is a parameter describing certain types of heavy new physics
(see Section \ref{stusec}).  In
addition, there are possible tree-level physics such as heavy $Z'$ bosons
or mixing with heavy exotic doublet leptons, $E'_R$, which could
significantly affect the asymmetry.  However, new physics probably cannot
explain all of the discrepancy with the other observables, because
some of the LEP observables measure precisely the same combination of
couplings as does $A_{LR}$\footnote{The relation makes use only of the
assumption that the LEP and SLD observables are dominated by the $Z$-pole.
The one loophole is the possibility of an important
contribution from other sources, such as new 4-fermi operators.  These are
mainly significant slightly away from the pole (at the pole they are out of
phase with the $Z$ amplitude and do not interfere). However, a combined
analysis of all constraints renders this possibility unlikely~\cite{jens}.}.
In particular, from the LEP
measurements of $A^{0\ell}_{FB}$, $A_e^0(P_\tau)$, and $A_\tau^0(P_\tau)$
one can obtain an average $A_{\ell \rm LEP}^0 = 0.147 \pm 0.004$,
consistent with the standard model prediction but 1.5$\sigma$ below $A_{LR}^0$.
If one does not assume lepton family universality, the LEP observables
$A^{0e}_{FB}$ and $A_e^0(P_\tau)$ imply
$A_{e \rm LEP}^0 = 0.141 \pm 0.007$, 1.7$\sigma$ below $A_{LR}^0$.

Finally, the forward-backward asymmetry into $\tau$'s,
$A_{FB}^{0\tau} = 0.0206 \pm 0.0023$ is 2.2$\sigma$ above
the standard model prediction, and 1.6$\sigma$ above the
average $0.0162 \pm 0.0014$
of $A_{FB}^{0\mu}$ and $A_{FB}^{0e}$. This is small enough
to be a fluctuation, so we will assume lepton flavor universality.

There are many other precision  observables.  Some recent ones
are shown in Table~\ref{tab2}.  These include the
$W$ mass
from CDF, D0, and UA2~\cite{a3},
the effective weak charge $Q_W$ measured in atomic parity violation
in cesium~\cite{cesium},
recent results on  the effective vector and axial couplings measured in
neutrino
electron scattering from CHARM II~\cite{a5}, and measurements of $s_W^2
\equiv 1 - M^2_W/M^2_Z$ from the CCFR collaboration at Fermilab~\cite{a6}.
This on-shell definition of the weak angle is determined from deep
inelastic neutrino scattering with small sensitivity to the top quark mass.
The result combined with earlier experiments~\cite{a7} is also shown.  All
of these quantities are in excellent agreement with the standard model
predictions.
\begin{table} \centering 
\begin{tabular}{|ccc|}  \hline 
Quantity & Value & Standard Model \\ \hline
$M_W$ (GeV) & $80.26 \pm 0.16$ & $80.34 \pm 0.01 \pm 0.04$ \\
$Q_W (C_S)$ & $-71.04 \pm 1.58 \pm [0.88]$ & $-72.88 \pm 0.05 \pm
0.03$ \\
$g_A^{\nu e}$ (CHARM II) & $-0.503 \pm 0.017$ & $-0.507 \pm 0 \pm
0.0004$ \\
$g_V^{\nu e}$ (CHARM II) & $-0.035 \pm 0.017$ & $-0.037 \pm 0.0005
\pm 0.0003$ \\
$s^2_W \equiv 1 - \frac{M_W^2}{M_Z^2}$ & $\begin{array}{c} 0.2218
\pm 0.0059 \;{\rm [CCFR]} \\ 0.2260 \pm 0.0048 \; {\rm [All]}
\end{array}$ & $0.2237 \pm 0.0002\pm 0.0008$ \\
$M_H$ (GeV) & $\geq 60$ LEP & $< \left\{ \begin{array}{c} 0
(600), \; {\rm theory} \\ 0 (800), \; {\rm indirect} \end{array}
\right.$ \\
$m_t$ & $ 180 \pm 12 $ CDF/D0 & $179 \pm 8 ^{+17}_{-20}$ [indirect]
\\ \hline
\end{tabular}
\caption{Recent observables from the $W$ mass and other
non-$Z$-pole observations compared with the standard model
expectations.  Direct limits and values on $M_H$ and $m_t$
are also shown.}
\label{tab2}
\end{table}

In the global fits to be described, all of the earlier low energy
observables not listed in the table are fully incorporated.
The electroweak corrections are now quite important.  The results
presented include full
1-loop corrections, as well as  dominant 2-loop effects, QCD corrections,
and mixed QCD-electroweak corrections.
For the renormalized weak angle, I use
the modified minimal subtraction
($\overline{MS}$) definition~\cite{a8}
$\sin^2 \hat{\theta}_W(M_Z) \equiv \hat{s}^2_Z$.
This basically means that one removes the
$\frac{1}{n-4}$ poles and some associated constants from the gauge
couplings. Other definitions are reviewed in~\cite{pdg95,el1}.The values
of $\hat{s}^2_Z$ obtained from various observables as a
function of $m_t$ are shown in figure \ref{fig1b}.
\begin{figure}
\postbb{30 310 525 675}{/home/pgl/fort/nc/graph/mt/xxmt.ps}{0.8}
\caption{Values of $\sin^2 \hat{\theta}_W (M_Z)$ as a function of
$m_t$ from various observables.}
\label{fig1b}
\end{figure}

\section{Results: $m_t$, $M_H$, $\alpha_s$, $\sin^2\theta_W$}
\label{parameters}

There are now sufficiently many observables that one can
determine $\hat{s}^2_Z$, $m_t$, and $\alpha_s (M_Z)$
from the $Z$-pole and other indirect precision data
simultaneously.  For example, $\hat{s}^2_Z$ can be determined
from the asymmetries,  $m_t$ from the $W$ and $Z$ masses, and
$\alpha_s (M_Z)$ from the hadronic $Z$-widths.  In practice all
of these quantities are determined from a simultaneous fit.  The
results of fits to various sets of data are shown in
Table~\ref{tab4}.
\begin{table}[t] \centering
\begin{tabular}{|ccccc|}  \hline 
Set & $\hat{s}^2_Z$ & $\alpha_s (M_Z)$ & $m_t$ (GeV) & $\Delta
\chi^2_H$ \\ \hline
Indirect $+$ CDF $+$ D0   & $0.2315 (2)(3)$ & $0.123
(4)(2)$ & $180\pm 7^{+12}_{-13}      $ & 7.9 \\
Indirect & $0.2315 (2)(2)$ & $0.123 (4)(2)$ & $179 \pm 8
^{+17}_{-20}$ & 8.5 \\
LEP  & $0.2318 (3)(2)$ & $0.124 (4)(2)$ &
$171 \pm 10^{+18}_{-20}$ & 5.7 \\
$Z$-pole (LEP $+$ SLD) & $0.2314 (3)(1)$ & $0.123 (4)(2)$ &
$181^{+8 \; +18}_{-9 \; -20}$ & 8.3 \\
SLD $+ \; M_Z$      & $0.2302 (5)(0)$ & ---            &
$220^{+14 \; +19}_{-15 \; -24}$ & \   \\ \hline
\end{tabular}
\caption{Results for the electroweak parameters in the standard model from
various sets of data.  The central values assume $M_H = 300$~GeV, while the
second errors are for $M_H \ra 1000 (+)$ and $60(-)$.  The last column is
the increase in the overall $\chi^2$ of the fit as $M_H$ increases from 60
to 1000.}
\label{tab4}
\end{table}
The first row of the table shows the global fit to all data,
including the direct production constraint $m_t = 180 \pm 12$ GeV
from CDF and D0. The second row uses the indirect data only.
The predicted value of $m_t = 179 \pm 8 ^{+17}_{-20}$ GeV
is in remarkable agreement with the CDF/D0 value.
The other fits show the sensitivity to the various data sets.

The LEP data allows a determination of the strong
coupling constant $\alpha_s$ at the $Z$-pole with a small experimental
error,
\beq \alpha_s (M_Z) = 0.123 \pm 0.004 \pm 0.002 \;\;\;\; {\rm
(lineshape)},\label{eq30a} \eeq
where the second uncertainty is from $M_H$. $\alpha_s$ is almost
uncorrelated with the other parameters.  It is determined mainly from the
ratio $R \equiv \Gamma (\rm had)/ \Gamma (\ell \bar{\ell})$, which is
insensitive to $m_t$ (except in the $b \bar{b}$ vertex), and also from
$\Gamma_Z$.  This determination is very clean theoretically, at least
within the standard model.  It is the $Z$-pole version of the long held
view that the ratio of hadronic to leptonic rates in $e^+e^-$ would be a
``gold plated'' extraction of $\alpha_s$ and test of QCD.  Using a recent
estimate \cite{a21} of the $(\alpha_s/\pi)^4$ corrections to $C_F$, {\it
i.e.} $- 90 (\alpha_s/\pi)^4$, one can estimate that higher-order terms
lead to an additional uncertainty $\sim \pm 0.001$ in the $\alpha_s(M_Z)$
value in (\ref{eq30a}).  It should be cautioned, however, that the
lineshape value is rather sensitive to the presence of some types of new
physics which affect the $Z$-hadron width, as is discussed below.

The lineshape value of $\alpha_s$ is in excellent agreement with the
independent value $\alpha_s (M_Z) = 0.123 \pm 0.006$ extracted from jet
event shapes at LEP using resummed QCD \cite{a22}.  It is also in
agreement with the prediction
\beq \alpha_s (M_Z) \sim 0.130 \pm 0.010 ,\;\;\;\;\; {\rm
SUSY-GUT} \eeq
of supersymmetric grand unification\footnote{In this case one should
actually use the value  $\alpha_s (M_Z) = 0.121(4)(1)$ appropriate
to the lower Higgs range expected in the supersymmetric extension
of the standard model.}~\cite{a27,lp4}.
As can be seen in Table~\ref{tab5},
however, it is somewhat larger than some of the low energy determinations
of $\alpha_s$ (which are then extrapolated theoretically to the $Z$-pole),
in particular those from deep inelastic scattering and the lattice
calculations of the charmonium and bottomonium spectra.
This slight discrepancy has led
some authors to suggest that there might be a light gluino which would
modify the running of $\alpha_s$,
or that there is a problem in the high energy determinations~\cite{shif}.
I think, however, that it is premature
to draw such  strong conclusions, especially since most of the
determinations are dominated by theoretical uncertainties.

There is, however, one significant uncertainty in the
lineshape value: if the high
experimental value of $R_b$ is due to a new physics contribution
to the $Z \rightarrow b \bar{b}$ vertex, and not just a fluctuation,
then the formulae for $R$ and $\Gamma_Z$ are affected, and the value
of $\alpha_s(M_Z)$ extracted from the lineshape is reduced~\cite{el1}.
Allowing for that possibility, one finds the lower  value
$\alpha_s(M_Z) = 0.101 \pm 0.008$, in better agreement with some
of the low energy determinations.

One could also consider the possibility that the low value of $R_c$
is due to new physics. However, allowing for that possibility, one obtains
$\alpha_s(M_Z) = 0.19 \pm 0.03$ or $0.16 \pm 0.04$, where the
former (latter) value does not (does) allow for new physics
in $R_b$ as well.  The first value in particular is in clear
disagreement with other determinations, so I will take the
view that $R_c$ is a statistical fluctuation.
\begin{table}                       \centering
\begin{tabular}{|lc|} \hline 
Source & $\alpha_s (M_Z)$ \\ \hline
$R_\tau$ & $0.122 \pm 0.005$ \\
Deep inelastic & $0.112 \pm 0.005$ \\
$\Upsilon$, $J/\Psi$ & $ 0.113 \pm 0.006$ \\
$c \bar{c}$ spectrum (lattice) & $0.110 \pm 0.006$ \\
$b \bar{b}$ spectrum (lattice) & $0.115 \pm 0.002$ \\
LEP, lineshape & $0.123 \pm 0.004 $ \\
LEP, event topologies & $0.123 \pm 0.006$ \\ \hline
\end{tabular}
\caption{Values of $\alpha_s$ at the $Z$-pole extracted from
various methods.}
\label{tab5}
\end{table}

\subsection{The Higgs Mass}

The new data significantly constrain  the Higgs boson mass.  This enters
the relation between $M_Z$ and the other observables
logarithmically and is strongly correlated with the
quadratic $m_t$   dependence in everything but  the $Z \ra b \bar{b}$
vertex correction.
The data
strongly favor a Higgs mass near the direct lower limit of $\sim$ 60 GeV.
This can be seen in the last column of Table~\ref{tab4}, which lists
the increase in $\chi^2$ for the overall fits as $M_H$ increases
from 60 to 1000 GeV. For example, in the fit to all data (including
the direct constraints) the
best fit is for $M_H$ = 60 GeV, with the limit $M_H < 320(430)$ GeV at
90(95)\% CL.
The allowed regions in $m_t$ vs. $M_H$ at various confidence
levels are shown in Figure~\ref{fig2}.

These low values for $M_H$ are consistent with the minimal
supersymmetric extension of the standard model, which generally predicts a
relatively light standard model-like Higgs scalar.  However,
a strong caveat is in order: the preference for small $M_H$ is
driven almost entirely by $R_b$ and $A_{LR}^0$, both
of which differ significantly from
the standard model predictions.
If these are due to
large statistical fluctuations or to some new physics then the constraint on
$M_H$ would essentially disappear.
\begin{figure}
\postbb{30 200 540 700}{/home/pgl/fort/nc/graph/mtmh/xxmhfit.ps}{0.6}
\caption{Allowed regions in $m_t$ vs. $M_H$ at various confidence
levels. The direct constraint $M_H > $ 60 GeV is also indicated.}
\label{fig2}
\end{figure}

The weak $M_H$ dependence does not imply that the data is insensitive to
the spontaneous symmetry breaking mechanisms.  Alternative schemes
generally yield large effects on the precision observables,
as will be described below.

\section{New Physics}
\subsection{Supersymmetry and Precision Experiments}

Let us now consider how the predictions for the precision observables are
modified in the presence of supersymmetry.  There are basically three
implications for the precision results.  The first, and most important, is
in the Higgs sector.  In the standard model the Higgs mass is arbitrary.  It
is controlled by an arbitrary quartic Higgs coupling, so that $M_H$ could
be as small as 60 GeV (the experimental limit) or as heavy as a TeV.  The
upper bound is not rigorous: larger values of $M_H$ would correspond to
such large quartic couplings that perturbation theory would break down.
This cannot be excluded, but would lead to a theory that is qualitatively
different from the (perturbative) standard model.  In particular, there are
fairly convincing triviality arguments, related to the running of the
quartic coupling, which exclude a Higgs which acts like a distinct elementary
particle for $M_H$ above $O(600$~GeV) \cite{a25}.

However, in supersymmetric extensions of the standard model the quartic
coupling is no longer a free parameter.  It is given by the squares of
gauge couplings, with the result that all supersymmetric models have at
least one Higgs scalar that is relatively light, typically with a mass
similar to the $Z$ mass.  In the minimal supersymmetric standard model
(MSSM) one has $M_H < 150$~GeV\footnote{At tree-level,
$M_H < M_Z$.}, which generally acts just like the standard
model Higgs\footnote{This is true if the second Higgs doublet is much
heavier than $M_Z$.} except that it is necessarily light.

In the standard model there is a large $m_t - M_H$ correlation, and
the global fit yields
\beq m_t \sim 180 \pm 7 + 13 \ln \left( \frac{M_H}{300 \rm
GeV} \right).\eeq
We have seen that for $60< M_H < 1000$~GeV this corresponds to
\beq m_t = 180 \pm 7 ^{+12}_{-13}\  ({\rm SM}). \eeq
However, in the MSSM one has the smaller range $60 < M_H < 150$~GeV,
leading to
\beq m_t = 169 \pm 7^{+4}_{-3} \ (\rm MSSM),\eeq
which is on the lower side of the CDF/D0 range, $(180 \pm 12$~GeV).
Because of the lower $m_t$, one obtains $\hat{s}^2_Z= 0.2313(2)(1)$
and $\alpha_s (M_Z) = 0.121 \pm 0.004^{+0.001}_{-0}$, which
differ slightly from the values in Table~\ref{tab4}.

There can be additional effects on the radiative corrections due to
sparticles and the second Higgs doublet that must be present in the MSSM.
However, for most of the allowed parameter space one has $M_{\rm new}
\gg M_Z$, and
the effects are negligible by the decoupling theorem.  For example, a large
$\tilde{t} - \tilde{b}$ splitting would contribute to the $\rho_0$
($SU_2$-breaking) parameter (to be discussed below), leading to a smaller
prediction for $m_t$, but these effects are negligible for $m_{\tilde{q}}
\gg M_Z$.  Similarly, there would be new contributions to the $Z\ra
b\bar{b}$ vertex for $m_{\chi^\pm}$, $m_{\tilde{t}}$, or $M_H^\pm \sim M_Z$.
The MSSM yields a better fit to the precision data than the standard
model~\cite{susyfits}, but that is mainly due to the
anomalous experimental value of $R_b$.

There are only small windows of allowed parameter space for which the new
particles contribute significantly to the  radiative corrections.  Except
for these, the only implications of supersymmetry from the precision
observables are: (a) there is a light standard model-like Higgs, which in
turn favors a smaller value of $m_t$.  Of course, if a light Higgs were
observed it would be consistent with supersymmetry but would not by itself
establish it.  That would require the direct discovery of the
superpartners, probably at the LHC.  (b) Another important implication of
supersymmetry, at least in the minimal model, is the {\it absence} of other
deviations from the standard model predictions.  (c) In supersymmetric
grand unification one expects the gauge coupling constants to unify when
extrapolated from their low energy values \cite{a27}.  This is consistent
with the data in the MSSM but not in the ordinary standard model (unless
other new particles or thresholds are added).  This is not actually a
modification of the
precision experiments, but a prediction for the observed gauge couplings.
Of course, one could have supersymmetry without grand unification.

\subsection{Unification of Gauge Couplings}
It is now well known that  the (properly normalized)
observed gauge couplings do not unify when
extrapolated to a large scale using the standard model
predictions for the running, but they do within experimental
uncertainties if they run according to the MSSM~\cite{a27}. Since
the electroweak couplings $\alpha(M_Z)$ and $\hat{s}^2_Z$ are known
precisely, it is useful to use them as inputs to
predict the more uncertain $\alpha_s$~\cite{lp4}. Using present data,
one predicts $\alpha_s(M_Z) = 0.130 \pm 0.001 \pm 0.010$,
where the first uncertainty is from the input couplings and the
second is an estimate of the theoretical uncertainties from the
low energy (supersymmetry) and superheavy (grand unification)
thresholds, and from possible nonrenormalizble operators (NRO).
As discussed
in Section~\ref{parameters}, this is in good agreement with
the experimental values determined from the $Z$ lineshape and
from the LEP jet event analysis, but is  high compared
with some low energy detrminations of $\alpha_s(M_Z)$. In contrast,
the non-supersymmetric standard model prediction is
$0.073 \pm 0.001 \pm 0.001$, well below the experimental values.

Thus, the observed couplings are consistent  to first approximation
with simple supersymmetric
grand unification and the associated concept of a grand desert
between the TeV and GUT scales. However:
\begin{itemize}
 \item Threshold corrections associated with the supersymmetric
particles are $> -0.003$ and are usually positive assuming
universal soft supersymmetry breaking terms~\cite{universal},
and supersymmetric contributions to the electroweak radiative
corrections generally lead to larger  $\alpha_s(M_Z)$~\cite{universal}.
Thus, low-scale threshold effects are not likely by
themselves\footnote{Non-universal gaugino masses can
lead to larger effects~\cite{sr}, but these would explicity
break the grand unification gauge symmetry.}
to lower the predicted $\alpha_s(M_Z)$ much below $\sim 0.125$.
The experimental value of  $\alpha_s(M_Z)$ is not a settled issue,
but if the lower values (e.g., 0.110-0.115) suggested by
some determinations are correct, one would have to invoke large but not
unreasonable GUT or string threshold effects, NRO, or intermediate
scale matter to maintain consistency.
 \item
The MSSM couplings unify at a scale $M_X \sim 3 \times 10^{16}$
GeV. This is far enough below the Planck scale $\sim 10^{19}$ GeV
that it may be consistent to consider grand unification of the strong
and electroweak couplings without gravity. Nevertheless, it is tempting
to bring gravity into the game. For example, one expects that
in superstring compactifications the couplings will unify around
the string scale $M_{\rm string} \sim g \times 5 \times 10^{17}$ GeV,
which is one order of magnitude above $M_X$. It is possible that the
string compactification first produces a grand unified theory in four
dimensions, which then breaks at the lower scale $M_X$. However,
it is difficult to find models in which this occurs and for which
the necessary Higgs multiplets to break the GUT symmetry are present.
Alternatively, the string compactification may lead directly to the
standard model group, in which case one must invoke string threshold
effects, intermediate scale matter, or higher
Ka\v c-Moody levels to explain the discrepancy of scales.
 \item Much attention has been focussed on the deviation of $M_X$ from
$M_{\rm string}$ and on the predicted $\alpha_s(M_Z)$.
However, the actual unification predictions are for $\ln(M_X/M_Z)$ and
$1/\alpha_s(M_Z)$. The former is  consistent with the string
scale to within 10\% and the latter is accurate
to within 15\%as well. Given the enormous number of
perturbations on the predictions
that can occur in string and GUT models one should view the predictions
as a major success for the general idea of GUT or string unification.
 \item
Most types of new physics at the TeV scale would have
very large (order 1) effects on the gauge unification prediction.
Unless the success  is just an accident,
and barring cancellations between large effects, this severely
restricts the possibilities for new physics beyond the MSSM
at the TeV scale. Essentially the only possibilities are
extended gauge groups which commute with the standard model
group, such as additional $U_1$ factors associated with
heavy $Z'$ bosons, complete ordinary or mirror additional
families, new exotic families which correspond to complete
GUT multiplets, and standard model singlets.
\end{itemize}

\subsection{Extended Technicolor/Compositeness}

In contrast, the other major class  of extensions, which includes
compositeness and dynamical symmetry breaking, leads to many implications
at low energies.  The most important are large flavor changing neutral
currents (FCNC).  Even if these are somehow evaded one generally expects
anomalous contributions to the $Z \ra b\bar{b}$ vertex, typically
$\Gamma(b\bar{b}) < \Gamma^{SM}(b\bar{b})$ in the simplest extended
technicolor (ETC) models \cite{a28}.  Similarly, one expects $\rho_0 \neq
1$, and $S \neq 0, T \neq 0$, where $\rho_0$, $S$,
and $T$ parameterize certain types of new physics, as will
be described below.  Finally, in theories with composite fermions one
generally expects new 4-fermi operators generated by constituent
interchange, leading to effective interactions of the form
\beq L = \pm \frac{4 \pi}{\Lambda^2} \bar{f}_1 \Gamma f_2 \bar{f}_3
\hat{\Gamma}f_4.\eeq
Generally, the $Z$-pole observables are not sensitive to such operators,
since they only measure the properties of the $Z$ and its
couplings\footnote{At the $Z$-pole the effects of new operators are out
of phase with the $Z$ amplitude and do not interfere. Interference
effects can survive away from the pole, but there
the $Z$ amplitude is smaller.}.
However, low energy experiments are sensitive.  In particular, FCNC
constraints typically set limits of order $\Lambda \geq O(100\; {\rm TeV})$
on the scale of the operators unless the flavor-changing effects are
fine-tuned away.  Even then there are significant limits from other flavor
conserving observables.  For example, atomic parity violation \cite{a29} is
sensitive to operators such as \cite{a30}
\beq L = \pm \frac{4\pi}{\Lambda^2} \bar{e}_L \gamma_\mu e_L
\bar{q}_L \gamma^\mu q_L .\eeq
The existing data already sets limits $\Lambda > O(10$~TeV).  Future
experiments should be sensitive to $\sim$ 40~TeV.

\subsection{The $Zb\bar{b}$ Vertex}

The $Zb\bar{b}$ vertex is especially interesting, both in the standard model
and in the presence of new physics.  In the standard model there are
special vertex contributions which depend quadratically on the top quark
mass. 
$\Gamma (b\bar{b})$
actually decreases with $m_t$~\cite{a18,a19},
as opposed to other widths which all
increase.  
The $m_t$ and $M_H$ dependences in the radiative corrections 
are strongly correlated, but the special vertex
corrections to $\Gamma(b\bar{b})$ are independent of $M_H$, allowing a
separation of $m_t$ and $M_H$ effects.

The vertex is also sensitive to a number of types of new physics.
One can parameterize such effects by \cite{a30a}
\beq \Gamma {(b\bar{b})}  \ra  \Gamma^{SM} {(b\bar{b})} \left( 1 +
\delta^{\rm new}_{bb} \right). \eeq
If the new physics gives similar contributions to vector and axial vector
vertices then the effects on $A_{\rm FB}^{b}$ are small.  In
supersymmetry one can have both positive and negative contributions
\cite{a31}.  In particular, light $\tilde{t} - \chi^{\pm}$ can give
$\delta^{\rm SUSY}_{bb} > 0$, as is suggested by the data, while light
charged Higgs particles can yield $\delta^{\rm Higgs}_{bb} < 0$.  In
practice, both effects are too small to be important in most allowed
regions of parameter space.  In extended technicolor (ETC) models there are
typically new vertex contributions generated by the same ETC interactions
which are needed to generate the large top quark mass.  It has
been argued that these are typically large and negative \cite{a28},
\beq \delta^{\rm ETC}_{bb} \sim - 0.056 \xi^2 \left(
\frac{m_t}{150{\rm GeV}} \right),\eeq
where $ \xi$ is a model dependent parameter of order unity.
They may be smaller in models with walking technicolor, but nevertheless
are expected to be negative and significant \cite{a33}.  This is in
contrast to the data, which suggests a positive contribution if any,
implying a serious problem for many ETC models. Possible ways out are
models in which the ETC and electroweak groups do not commute, for which
either sign is possible \cite{a34}, models in which diagonal interactions
related to extended technicolor dominate~\cite{etcdiag}, and topcolor
or topcolor assisted technicolor models, for which the ETC contribution
to $m_t$ is small~\cite{topcolor}.

Another possibility is mixing  between the $b$ and exotic heavy fermions
with non-canonical weak interaction quantum numbers.  Many extensions of
the standard model predict, for example, the existence of a heavy
$D_L$, $D_R$, which are both $SU_2$ singlet quarks with charge $-1/3$.
These can mix with the $d$,
$s$, or $b$ quarks, but one typically expects such mixing to be largest for
the third generation.  However, this mechanism gives a negative
contribution
\beq \delta_{bb}^{D_L} \sim -2.3 s^2_L \eeq
to $\delta^{\rm new}_{bb}$, where $s_L$ is the sine of the $b_l - D_L$
mixing angle.

One can extract $\delta^{\rm new}_{bb}$ from the data, in a global fit to
the standard model parameters as well as $\delta^{\rm new}_{bb}$.  This
yields
\beq \delta^{\rm new}_{bb} = 0.032 \pm 0.010,  \eeq
which is $\sim 3.2\sigma$ above zero.  This value is hardly changed when
one allows additional new physics, such as described by the $S$, $T$, and
$U$ parameters.  $\delta^{\rm new}_{bb}$ is correlated with
$\alpha_s(M_Z)$: one obtains $\alpha_s(M_Z) = 0.101 \pm 0.008$,
considerably smaller than the standard model value $0.123 (4)(2)$. Allowing
$\delta_{bb}^{\rm new} \neq 0$ has negligible effect on $\hat{s}^2_Z$ or
$m_t$.

In a more detailed analysis~\cite{el1}
one can allow separate corrections in the left and right-handed $b$
couplings,
\beq g_{Lb} = \frac{1}{2} ( g_{Vb} + g_{Ab}) \RA -\frac{1}{2} +
\frac{1}{3} \sin^2 \theta_W + \delta^b_L, \eeq
and
\beq g_{Rb} = \frac{1}{2} ( g_{Vb} - g_{Ab}) \RA
\frac{1}{3} \sin^2 \theta_W + \delta^b_R. \eeq
Then $R_b$, $A_{FB}^{0b}$ and the other observables can be used
to simultaneously constrain $\delta^b_L$, $\delta^b_R$, and the
other parameters. From the global fit
\beq
\delta^b_L = -0.0033 \pm 0.0035 \ \ \ \ \ \ \
\delta^b_R = 0.018 \pm 0.013, \eeq
with a correlation of 0.80. These should be compared with the
unperturbed standard model values  $g_{Lb}=-0.421$ and
$g_{Rb}=0.077$, respectively. One also obtains the low value
$\alpha_s(M_Z) = 0.101 \pm 0.008$, just as in the single parameter case.
At their central values
$\delta^b_L$ and $\delta^b_R$ contribute about equally to
$R_b$, while their effects partially cancel in $A_{FB}^{0b}$,
which is consistent with the standard model.
The allowed region in $\delta^b_L - \delta^b_R$
is shown in Figure~\ref{blr}. One sees that there is a tendency from the
anomaly to be in $\delta^b_R$, but it cannot be excluded that the
effect is in $\delta^b_L$ or a mixture.

\begin{figure}
\postbb{60 240 535 690}{/home/pgl/fort/nc/graph/st/xxzbb.ps}{0.6}
\caption{Allowed regions in
$\delta^b_L$ vs $\delta^b_R$ for $M_H = 300$ GeV.}
\label{blr}
\end{figure}

\subsection{$\rho_0$: Nonstandard Higgs  or Non-degenerate Heavy
Multiplets}

One parameterization of certain new types of physics is the parameter
$\rho_0$, which is introduced to describe new sources of $SU_2$ breaking
other than the ordinary Higgs doublets or the top/bottom splitting.
One defines $\rho_0 \equiv M_W^2/(M_Z^2 \hat{c}^2_Z \hat{\rho})$,
where $\hat{c}^2_Z \equiv 1 - \hat{s}^2_Z$;
$\hat{\rho} \sim 1 + 3 G_F m_t^2 /8 \sqrt{2} \pi^2 $
absorbs the relevant standard model radiative
corrections so that $\rho_0 \equiv 1$ in the standard model.
New physics can affect $\rho_0$ at either the tree or loop-level
\beq \rho_0 = \rho_0^{\rm tree} + \rho_0^{\rm loop}.\eeq
The tree-level contribution is given by Higgs representations
larger than doublets, namely,
\beq \rho_0^{\rm tree} = \frac{\sum_i \left( t^2_i -  t_{3i}^2
+ t_i \right) |\langle \phi_i \rangle|^2}{ \sum_i 2 t_{3i}^2
|\langle \phi_i \rangle|^2},  \label{eqerica}\eeq
where $t_i$ ($t_{3i}$) is the weak isospin (third component) of the
neutral Higgs field $\phi_i$.
For Higgs singlets and doublets ($t_i = 0,\frac{1}{2}$) only,
$\rho_0^{\rm tree} = 1$.
However, in the presence of larger representations with non-zero
vacuum expectation values
\beq \rho_0^{\rm tree} \simeq 1 + 2 \sum_i \left( t^2_i -
3 t_{3i}^2 + t_i \right) \frac{ |\langle \phi_i \rangle
|^2}{|\langle \phi_{\frac{1}{2}} \rangle |^2 }. \label{eq22a3}
\eeq

One can also have loop-induced contributions similar to that of the
top/bottom, due to non-degenerate multiplets of fermions or bosons.  For new
doublets
\beq \rho_0^{\rm loop} = \frac{3G_f}{8 \sqrt{2} \pi^2} \sum_i
\frac{C_i}{3} F (m_{1i},m_{2i}),   \eeq
where $C_i = 3(1)$ for color triplets (singlets) and
\beqa F(m_1, m_2) &  = & m_1^2 + m^2_2 - \frac{4m_1^2 \; m^2_2}{m_1^2 -
m^2_2} \ln \frac{m_2}{m_2} \nonumber \\ & \geq  & (m_1- m_2)^2 . \eeqa
Loop contributions to $\rho_0$ are generally positive,\footnote{One can
have $\rho^{\rm loop} < 0$ for Majorana fermions \protect\cite{a34a} or
boson multiplets with vacuum expectation values \protect\cite{a34b}.} and
if present would lead to lower values for the predicted $m_t$. $\rho_0^{\rm
tree} - 1$ can be either positive or negative depending on the quantum numbers
of the Higgs field.  The $\rho_0$ parameter is extremely important because
one expects $\rho_0 \sim 1$ in most superstring theories, which generally
do not have higher-dimensional Higgs representations, while typically
$\rho_0 \neq 1$ from many sources in models involving compositeness.

In the presence of $\rho_0$ the standard model formulas for the observables
are modified.  One has
\beq M_Z  \ra  \frac{1}{\sqrt{\rho_0}} M_Z^{SM}, \ \ \ \Gamma_Z \ra
\rho_0
\Gamma_Z^{SM}, \ \ \
{\cal{L}}_{NC}   \ra   \rho_0 {\cal{L}}^{SM}_{NC}.\eeq
It has long been known that $\rho_0$ is close to 1. However, until recently it
has been difficult to separate $\rho_0$ from $m_t$, because in most
observables one has only the combination $\rho_0 \hat{\rho}$.  The one
exception has been the $Z \ra b\bar{b}$ vertex.  However, the direct
observation of the $t$ by CDF and D0, with their average $m_t
= 180 \pm 12$ GeV, allows one to
calculate $\hat{\rho}$ and therefore separate $\rho_0$.  In practice one
fits to $m_t$, $\rho_0$ and the other parameters, using the CDF/D0 value
of $m_t$ as an additional constraint.  One can determine
$\hat{s}^2_Z$, $\rho_0$, $m_t$, and $\alpha_s$ simultaneously, yielding
\beqa
\rho_0 & = & 1.0012
\pm 0.0013 \pm 0.0018 \nonumber \\
\hat{s}^2_Z & = & 0.2314(2)(2)
\nonumber \\
 \alpha_s  & = & 0.121(4)(1) \nonumber \\  m_t & = & 171 \pm 12 \
{\rm GeV},  \label{eqrho} \eeqa
where the second uncertanty is from $M_H$.  Even in the presence of the
classes of new physics parameterized by $\rho_0$ one still has robust
predictions for the weak angle and a good determination of $\alpha_s$.
Most remarkably, given the CDF/D0 constraint, $\rho_0$ is constrained to be
very close to unity, causing serious problems
for compositeness models.  The allowed region in $\rho_0$ vs $\hat{s}^2_Z$ are
shown in Figure \ref{figerica}.  This places limits $|\langle \phi_i
\rangle|/ |\langle \phi_{1/2} \rangle| < {\rm few} \%$ on non-doublet
vacuum expectation values, and places constraints $\frac{C}{3} F(m_1, m_2)
\leq (100\; {\rm GeV})^2$ on the splittings of additional fermion or boson
multiplets.

\begin{figure}
\postbb{60 230 535 680}{/home/pgl/fort/nc/graph/xrho/xxrhof.ps}{0.6}
\caption{Allowed regions in $\rho_0$ vs $\hat{s}^2_Z$ for $M_H = 60$, 300,
and 1000 GeV.}
\label{figerica}
\end{figure}

\subsection{Heavy Physics by Gauge Self Energies}
\label{stusec}

A larger class of extensions of the standard model can be parameterized by
the $S$, $T$ and $U$ parameters \cite{a35}, which describe that subset of
new physics which affect only the gauge boson self-energies but do not
directly affect new vertices, etc.  One introduces three
parameters\footnote{Three additional parameters are needed
if the new physics scale is comparable to $M_Z$~\cite{sixpar}.}
\begin{eqnarray} S &=& S_{\rm new} + S_{m_t} + S_{M_H} \nonumber \\
                 T &=& T_{\rm new} + T_{m_t} + T_{M_H}  \\
   U &=& U_{\rm new} + U_{m_t}. \nonumber \end{eqnarray}
$S$ describes the breaking of the $SU_{2A}$ axial generators and is
generated, for example, by degenerate heavy chiral families of fermions.
$T$ and $U$ describe the breaking of $SU_{2V}$ vector generators:
 $T$ is equivalent to the
$\rho_0$ parameter and is induced by mass splitting in multiplets of
fermions or bosons. $U$ is zero in most extensions of the standard model. $S$,
$T$ and $U$ were introduced to describe the contributions of new physics.
However, they can also parametrize the effects of very heavy $m_t$ and
$M_H$ (compared to $M_Z$).  Until recently it was difficult to separate the
$m_t$ and new physics contributions.  Now, however, with the CDF/D0
measurement of
$m_t$ it is possible to directly extract the new physics contributions.
In the following, I will use $S$, $T$, and $U$ to represent the
effects of new physics only, with the $m_t$ and $M_H$ effects on
observables included separately.

A new multiplet of degenerate chiral fermions will contribute to $S$ by
\beq S |_{\rm degenerate} = C_i |t_{3L} (i) - t_{3R} (i)
|^2/3\pi \geq 0, \eeq
where $C_i$ is the number of colors and $t_{3LR}$ are the $t_3$ quantum
numbers.  A fourth family of degenerate fermions would yield
$\frac{2}{3\pi} \sim 0.21$, while QCD-like technicolor models, which
typically have many particles, can give larger contributions.  For example,
$S \sim 0.45$ from an isodoublet of fermions
with four technicolors, and an
entire technigeneration would yield $1.62$ \cite{a36}.  Non-QCD-like theories
such as those involving walking could yield smaller or even
negative contributions \cite{a37}.  Loops involving scalars,
nondegenerate fermions, or Majorana particles can
contribute to $S$ with either sign \cite{a38}.
(Note that $S$, $T$, and
$U$ are induced by loop corrections and have a factor of $\alpha$ extracted,
so they are expected to be $O(1)$ if there is new physics.)

The $T$ parameter is analogous to $\rho_0^{\rm loop}$.  For a
non-degenerate family
\beq T \sim \frac{\rho_0^{\rm loop}}{\alpha} \sim 0.42
\frac{\Delta m^2}{(100 \;GeV)^2}, \eeq
where
\beq \Delta m^2  =  \sum_i \frac{C_i}{3} F \left( m_{1i}, m_{2i} \right).
\eeq
Usually $T > 0$, although there may be exceptions for theories
with Majorana fermions or additional Higgs doublets.  In practice,
higher-dimensional Higgs multiplets could mimic $T$ with either
sign (see equation (\ref{eqerica})), and cannot be separated from loop
effects unless they are seen directly or have other effects.
I will therefore deviate slightly from the historical definition
and redefine $T$ to include the tree level effects of such
multiplets, so that
\beq \rho_0 = \frac{1}{1 - \alpha T} \simeq 1 + \alpha T. \eeq
 Usually $U$ is small.

There is enough data to simultaneously determine the new physics
contributions to $S$, $T$, and $U$, the standard model parameters, and also
$\delta^{\rm new}_{bb} =
\frac{\Gamma (b\bar{b})}{\Gamma^{\rm SM} (b\bar{b})} -1
$. For example, $S$, $T$, $U$, $\delta^{\rm
new}_{bb}$, $\hat{s}^2_{Z}$, $\alpha_s(M_Z)$ and $m_t$ are constrained by
$M_Z$, $\Gamma$, $M_W$, $R_b$, asymmetries, $R$, and $m_t$ (CDF/D0),
respectively.  One obtains
\begin{eqnarray} S & = & -0.28 \pm 0.19^{-0.08}_{+0.17} \nonumber \\
T & = & -0.20 \pm 0.26^{+0.17}_{-0.12} \nonumber \\
U &  = & -0.31 \pm 0.54 \nonumber \\
\delta^{\rm new}_{bb} & = & 0.032 \pm 0.010 \nonumber \\
 \hat{s}^2_Z & = & 0.2311 (3) \nonumber \\
\alpha_s(M_Z) & = & 0.103 (8) \nonumber \\
 m_t & = & 181 \pm 12 \; {\rm GeV},
\end{eqnarray}
where the second error is from $M_H$.
The $T$ value corresponds to
$\rho_0 = 0.9985 \pm 0.0019 ^{+ 0.0012}_{- 0.0009}$, which differs
from the value in (\ref{eqrho}) because of the presence of $S$,
$U$, and $\delta^{\rm new}_{bb}$.
The data is consistent with the
standard model: $S$ and $T$ are consistent with the standard model
expectation of 0 at or near the 1$\sigma$ level, although there is
a small tendency for negative values.
The constraints on $S$ are a problem for those
classes of new physics such as QCD-like technicolor which tend to give $S
$ large and positive, and $S$ allows, at most, one additional
family of ordinary  or mirror
fermions at 95\% CL.  (Of course the invisible $Z$ width
precludes any new families unless the additional neutrinos are heavier than
$M_Z/2$.)  The allowed regions in $S$ vs $T$ are shown in Figure~\ref{fig5}.
The seven parameter fit still favors a non-zero $Z\ra b\bar{b}$ vertex
correction $\delta^{\rm new}_{bb}$. As in the  model with $S=T=U=0$, the
extracted $\alpha_s(M_Z)$ is strongly correlated with
$\delta^{\rm new}_{bb}$. For $\delta^{\rm new}_{bb}=0$ one would obtain
$\alpha_s(M_Z)= 0.122(5)$.

\begin{figure}[tbh]
\postbb{60 230 520 670}{/home/pgl/fort/nc/graph/st/xxst.ps}{0.7}
\caption{Constraints on $S$ and $T$ from various
observables and from the global fit to all data. $S$ and $T$
represent the effects of new physics only; uncertainties from
$m_t$ are included in the errors.}
\label{fig5}
\end{figure}

\section{Conclusions}

\begin{itemize}
\item  The precision data have confirmed the standard electroweak model.
However, there are possible hints of discrepancies at the 2 -- 3 $\sigma$
level in $\Gamma (b\bar{b})/\Gamma(\rm had)$ and $A^0_{LR}$.

\item  The data not only probes the tree-level structure, but the
electroweak loops have been observed at the $2\sigma$ level.  These consist
of much larger fermionic pieces involving the top quark and QED, which only
partially cancel the bosonic loops.  The bosonic loops, which probe
non-abelian vertices and gauge-Higgs vertices, are definitely needed to
describe the data.

\item  The global fit to the data (including the constraint
$m_t = 180 \pm 12$ GeV from CDF and D0)
within the standard model yields
\beqa \overline{MS}: \hat{s}^2_Z & = & 0.2315 (2)(3) \nonumber \\
{\rm on-shell:} \; s^2_W
& \equiv & 1 - \frac{M_W^2}{M_Z^2}  = 0.2236 (8)  \nonumber \\
m_t &  = & 180 \pm 7\,^{+12}_{-13} \nonumber \\ \alpha_s (M_Z)
 & =  & 0.123 (4)(2),  \eeqa
where the second uncertainty is from $M_H$.
A fit to the indirect data only yields the prediction
$m_t = 179 \pm 8\,^{+17}_{-20}$
GeV, in remarkable agreement with the direct CDF/D0 value.
The data  also allow a clean and
precise extraction of $\alpha_s$ from the lineshape.  This is in excellent
agreement with the value $\alpha_s (M_Z) = 0.123 \pm 0.006$ from event
shapes.  Both are larger than many of the low energy determinations when
extrapolated to the $Z$-pole.  The lineshape determination, however, is
sensitive to the presence of certain types of new physics.

\item  The agreement between the indirect prediction for $m_t$
with the  direct CDF/D0 observation,
and of $\alpha_s$ with the various other determinations
is an impressive success for the
entire program of precision observables.

\item  Combining the direct CDF/D0 value of $m_t$ with the indirect constraints
does not make a large difference within the context of the standard model.
However, when one goes  beyond the standard model, the direct $m_t$
allows a clean extraction of the new physics contributions to $\rho_0$,
which is now shown to be very close to unity, $\rho_0 = 1.0012 (13)(18)$.
This strongly limits Higgs triplet vacuum expectation values and
non-degenerate heavy multiplets.  Similarly, it allows an extraction of the
new physics contributions to $S$, $T$, $U$,
which are consistent with zero.  Finally, one can determine the new physics
contributions to the $b\bar{b}$ vertex: $\delta^{\rm new}_{bb}$ is
approximately $3.2\sigma$ away from zero, reflecting the large value of the
$b\bar{b}$ width.

\item  The data exhibit a  preference for a light Higgs.
One finds  $M_H \leq 320 (420)$~GeV
at 90(95\%) CL.  However, the preference depends crucially on the
observed values of $R_b$ and $A^0_{LR}$, both of which differ
significantly from the standard model expectations. If these are
due to new physics, the $M_H$ constraint is relaxed or disappears.

\item  The major prediction of supersymmetry is that one does not expect
large deviations in the precision observables.  The new particles tend to
be heavy and decouple.  One implication that is relevant, however, is that
supersymmetric theories have a light standard model-like Higgs.  They
therefore favor the lighter Higgs mass and the lower end of the predicted
$m_t$ range.

\item  The observed gauge couplings are consistent within 15\% with the
coupling constant unification expected in supersymmetric grand unification,
but not with the simplest version of non-supersymmetric unification.
The logarithm of the unification scale is also consistent within 10\% with
the expectations of superstring compactifications which break directly
to the standard model group. Perhaps we should take this as a hint
that the grand desert hypothesis is correct, and focus on GUTs and
string compactifications for which threshold and new particle
effects are small (of order 10\%).

\item  In compositeness and dynamical symmetry breaking theories one
typically expects not only large flavor changing neutral currents but
significant deviations of $\rho_0$ from unity and of $S$ and
$T$ from zero.  One further expects that $\delta_{bb}^{\rm new} <
0$, at least in the simplest models.  Therefore, the precision experiments
are a major difficulty for this class of models.

\end{itemize}

\section*{Acknowledgement}

It is a pleasure to thank Jens Erler for collaboration on these analyses.
I would also like to thank the conference organizers for travel support.
This work was supported by U.S. Department of Energy Grant No.
DOE-EY-76-02-3071 and the National Science Foundation Grant No. PHY94-07194.


\end{document}